\documentclass[showkeys,preprint,showpacs,nopreprintnumbers,amsmath,amssymb,aps]{revtex4}

\usepackage{graphicx}
\usepackage{dcolumn}
\usepackage{bm}

\begin{document}

\pagenumbering{arabic}

\title{Single-crystalline nanopillars for spin-transfer 
measurements} 

\author{H. Dassow}
\author{R. Lehndorff}
\author{D.~E. B\"urgler}
\thanks{Corresponding author. Email: d.buergler@fz-juelich.de, phone: 
+49 2461 614 214, FAX: +49 2461 614 443}
\author{M.~Buchmeier}
\author{P.~A. Gr\"unberg}
\author{C.~M. Schneider}
\affiliation{Institute of Solid State Research - Electronic 
Properties (IFF6) and \\ cni -- Center of Nanoelectronic Systems for 
Information Technology, Research Center J\"ulich GmbH, D-52425 
J\"ulich, Germany}
\author{A.~van~der~Hart}
\affiliation{Institute of Thin Films and Interfaces - Process Technology 
(ISG-PT) and \\
cni -- Center of Nanoelectronic Systems for Information Technology, 
Research Center J\"ulich GmbH, D-52425 J\"ulich, Germany}

\date{\today}

\begin{abstract}
We report on current-induced magnetization switching (CIMS) in single- 
crystalline nanopillars.  
Fe(14\,nm)/Cr(0.9\,nm)/Fe(10\,nm)/Ag(6\,nm)/Fe(2\,nm) multilayers are 
deposited by molecular-beam epitaxy.  The central Fe layer is coupled 
to the thick one by interlayer exchange coupling over Cr.  The topmost 
Fe layer is decoupled (free layer).  Nanopillars with 150\,nm diameter 
are prepared by optical and e-beam lithography.  The opposite spin 
scattering asymmetries of Fe/Cr and Fe/Ag interfaces enable us to 
observe CIMS at small magnetic fields and opposite current polarity in 
a single device.  At high magnetic fields, step-like resistance 
changes are measured at positive currents and are attributed to 
current-driven magnetic excitations.
\end{abstract}

\pacs{75.47.-m, 72.25.Ba, 75.60.Jk}

\keywords{Single-Crystalline Nanopillar, Molecular Beam Epitaxy, 
Giant Magneto Resistance, Current Induced Magnetization Switching, 
Spin-Transfer Torque}

\maketitle

In a magnetic multilayer containing two ferromagnetic layers and a 
nonmagnetic spacer (FM2/NM/FM1), an electric current flowing 
perpendicularly to the layers (CPP) gets spin-polarized by the FM 
layers, leading to a giant magnetoresistance 
(GMR)\cite{baibich88,binasch89}.  Thus, spin currents can sense the 
magnetization state of the magnetic system.  Slonczewski 
\cite{slonczewski96} and Berger \cite{berger96} first predicted that 
spin currents of appropriate strength can also directly influence the magnetizations 
without applying an external magnetic field.  Electrons flowing from 
FM2 to FM1 are first polarized by FM2 and then repolarized at the 
interface NM/FM1, where the transverse component of the spin current 
is absorbed and acts as a torque on the magnetic moment $\bm{M_1}$ of 
FM1 \cite{stiles02}. By reversing the current direction, the spin 
current reflected from FM2 is repolarized at the NM/FM1 interface 
leading to a reversed torque.  Therefore, FM1 can be switched from the 
parallel to the antiparallel configuration with respect to FM2 back 
and forth by repeatedly reversing the current polarity, as long as 
$\bm{M_2}$ remains fixed.  This pinning can be achieved by different FM 
layer thicknesses \cite{albert00,grollier01,urazhdin03}, by the 
exchange bias effect \cite{krivorotov05}, or by making use of 
interlayer exchange coupling, as in our case.

In order to achieve large spin-torque effects a high spin polarization 
$P$ of the current is needed.  Thus, the present work is motivated by 
two publications of Stiles and Penn \cite{stiles00} and Stiles and 
Zangwill \cite{stiles02}, in which the authors predict high 
spin polarization for single-crystalline Fe/Ag interfaces.  
Single-crystalline layered structures can also serve as model systems 
for comparison with theory due to the well known structure, the small 
amount of defects, and the homogeneous magnetic properties when 
prepared by molecular beam epitaxy (MBE).  In particular, 
single-crystalline Fe(001) layers show well-defined 4-fold in-plane 
magnetocrystalline anisotropy, which also helps to stabilize the 
magnetization along two easy axes ([100] and [010]).


First, we deposit the magnetic multilayer.  In order to achieve 
single-crystalline growth we use a standard MBE system.  The native 
oxygen layer of the GaAs(001) substrates ($10\times 10\, 
\mathrm{mm}^2$) is desorbed by annealing for 60\,min at $580^\circ 
\mathrm{C}$ under UHV conditions.  We deposit 1\,nm Fe and 150\,nm Ag 
at $100^\circ \mathrm{C}$ to get a flat buffer system after annealing 
at $300^\circ \mathrm{C}$ for $1\,\mathrm{h}$ \cite{buergler96}.  The 
Ag buffer also act as a bottom electrode for the transport 
measurements.  The following layers are then deposited at room 
temperature: Fe(14)/Cr(0.9)/Fe(10)/Ag(6)/Fe(2) [thicknesses in nm].  
We check the crystalline surface structure after each deposited layer 
by low-energy electron diffraction (LEED).  The spots characteristic 
of (001) surfaces slightly broaden with increasing total 
thickness, but still indicate high crystalline quality, even for the 
final 50~nm Au(001) capping layer.  Thicknesses are controlled by quartz 
crystal monitors.  The bottom and central FM layers [Fe(14) and Fe(10)] are 
coupled by interlayer exchange coupling over the Cr interlayer.  
Therefore, the central Fe(10) layer is magnetically harder with respect to the 
top Fe(2) layer.

\begin{figure}
\includegraphics[width=8cm]{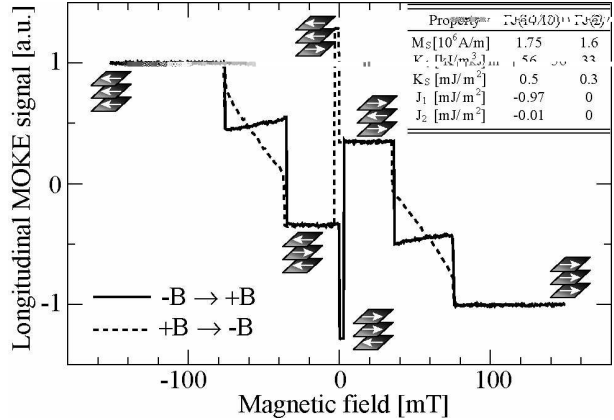} 
\caption{\label{fig:fig_1} MOKE hysteresis loop of the extended 
layered system measured with the external magnetic field parallel to 
one of the easy [100]-axes of Fe.  The interlayer exchange coupling 
stabilizes the fully antiferromagnetic state below $\pm 35\,$mT. The 
top Fe(2) is decoupled from the rest of the system as indicated by the 
negligible coupling constants $J_1$ and $J_2$ (see inset) of the Fe(2) 
layer, extracted from additional BLS measurements \cite{buchmeier05}.}
\end{figure}

The longitudinal magneto-optical Kerr effect (MOKE) is used to 
measure the magnetic properties of the samples. In 
Fig.~\ref{fig:fig_1} we present the hysteresis loop with the 
magnetic field parallel to one of the easy [100]-axes of the Fe 
layers in the film plane. The saturation field of the system is 
$|B_\mathrm{S}|=76\,\mathrm{mT}$. For smaller magnetic fields the 
central Fe(10) layer remagnetizes via a canted state to the fully 
antiferromagnetic configuration of the trilayer stack below the switching 
field $\pm 35\,$mT. After reversing the field direction we measure 
another jump in the signal, which corresponds to the reversal of the 
topmost $2\,$nm thick Fe layer at $\pm 0.3\,$mT. At $\pm 3\,$mT 
the two coupled Fe layers reverse simultaneously due to their unequal 
thickness. 

By fitting the MOKE measurements and additional Brillouin Light 
Scattering measurements (BLS) \cite{buchmeier03a} we extract the 
magnetic properties of each layer as compiled in the inset of 
Fig.~\ref{fig:fig_1}.  The saturation magnetization $M_\mathrm{S}$
and the crystalline anisotropy $K_1$ have bulk values 
\cite{kittel96,wohlfarth80} and indicate the high quality of the 
layers.  The thin Fe(2) layer has reduced $M_\mathrm{S}$ and $K_1$, 
which can be understood by the reduced thickness or by reduced growth 
quality.  The negligible coupling constants $J_1$ and $J_2$ show that 
this layer is decoupled.  $K_\mathrm{S}$ denotes the interface 
anisotropy.


In order to measure the spin-transfer effects in the CPP-geometry we 
have developed a combined process of optical and e-beam lithography.  
First, we define the leads and contact pads of the bottom electrode by 
using AZ5206 photoresist and ion beam etching (IBE).  We then employ 
HSQ (hydrogen silsesquioxane) as negative e-beam sensitive resist 
\cite{namatsu98} and a Leica EBPG 5HR e-beam writer to define small 
nanopillars.  The resist structures are circular and transferred into 
the magnetic layers by IBE. The timed etching process is stopped 
inside the magnetic multilayer.  Typical dimensions of the developed 
resist structures are 100\,-\,150\,nm (measured with an atomic 
force microscope).  Due to redeposition of etched material during IBE 
\cite{gloersen75}, the nanopillars broaden to 150\,-\,200\,nm.  
The pillars are planarized by spin-coating HSQ. Subsequent e-beam 
exposure turns HSQ into SiO$_\mathrm{x}$, which electrically insulates 
the pillars \cite{namatsu98}.  In order to improve the insulation, 
especially at the side walls of the bottom electrodes, a 50\,nm 
Si$_3$N$_4$ layer is deposited by plasma enhanced chemical vapor 
deposition (PECVD).  We open the top of the nanopillars by IBE and use 
an optical lift-off process of 300\,nm Au for the preparation of the 
top electrode for the 4-point resistance measurements.


The DC voltage drop of a constant current $I$ applied to the junction 
is measured, and by dividing by $I$ we calculate the absolute 
resistance $R$.  The differential resistance $dU/dI$ is recorded with 
lock-in technique by mixing a constant current with a small modulated 
voltage ($\approx 300\,\mu\mathrm{V}$ and $\approx 12\,$kHz). 
Typical junction resistances lie in the range between 1 and 3\,$\Omega 
$.  The temperature can be controlled with a He flow cryostat between 
4 and 300\,K.

\begin{figure}
\includegraphics[width=8cm]{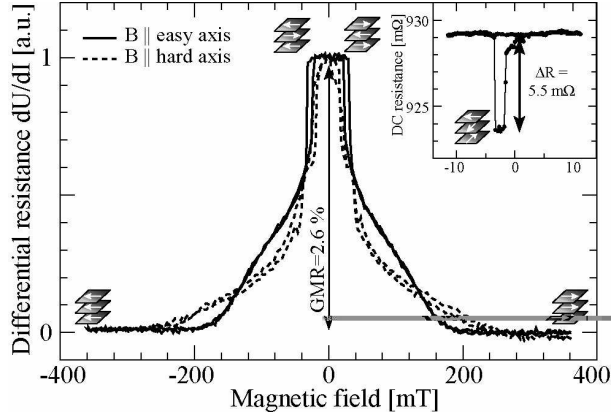} 
\caption{\label{fig:fig_2} GMR hysteresis loop with magnetic field 
parallel to an easy (solid) and hard (dashed) axis. 
 Inset: Minor GMR loop. Only in the first half of the 
loop ($+B\rightarrow -B$) the resistance drops to a smaller value 
corresponding to a canted magnetization state. These drops occur not 
in every cycle. In the second half the resistance stays at the 
maximum value.}
\end{figure}

\begin{figure}
\includegraphics[width=8cm]{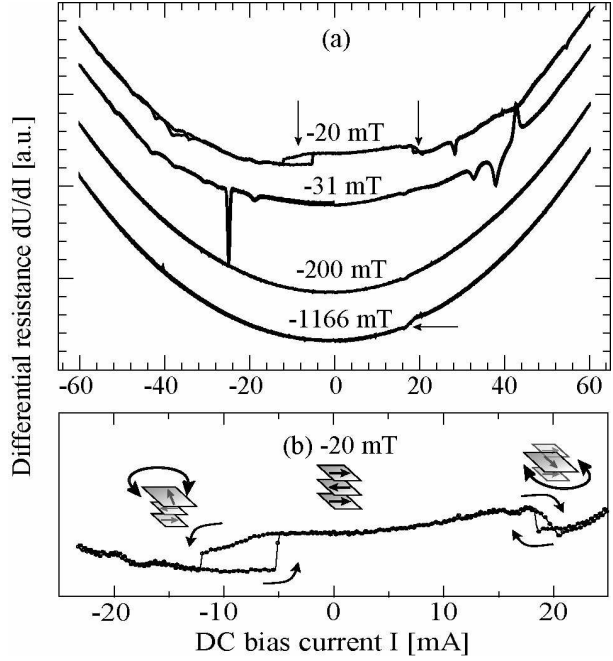} 
\caption{\label{fig:fig_3} (a) DC current loops with magnetic field 
parallel to an easy axis. Curves measured at different external fields 
as indicated are offset vertically for 
clarity.  
(b) Expanded view of the curve for -20\,mT with 
the interpretation of the switching processes occurring at positive 
and negative DC bias current.}
\end{figure}

The magnetoresistance loop of a junction without applying a DC bias 
current is shown in Fig.~\ref{fig:fig_2}.  The solid (dashed) line 
represents the data with magnetic field along the easy (hard) axis of 
Fe(001).  The curves show a completely different behavior for the two 
field directions, but are the same along the second pair of easy and hard 
axes.  Thus, the structure is still single-crystalline and exhibits 
4-fold magnetocrystalline anisotropy.  The saturation field of the 
structured sample is 190\,mT, which is more than twice the saturation 
field of the extended layers (see Fig.~\ref{fig:fig_1}).  Another 
difference becomes obvious in the minor loop (inset of 
Fig.~\ref{fig:fig_2}), where the absolute resistance is measured with 
a small DC current of 1\,mA. Coming from large positive magnetic 
field, the resistance drops to a smaller value at small reversed 
fields between 1 and 3\,mT and jumps back to the high resistance 
state at larger negative fields.  On the way back, the resistance 
stays at the maximum value.  The drop on the first half of the cycle 
does not occur on every measurement.  Thus, the patterning has 
modified the magnetic configuration and the structured Fe(2) 
nanomagnet is presumably coupled to the rest of the system by dipolar stray 
fields at the edges or by domain wall coupling.  This is a common 
feature in these devices also seen in Co nanopillars \cite{albert00}.  
Due to this effect, we cannot separate the contributions of both 
subsystems to the GMR, and therefore cannot gauge the resistance jumps 
measured under the influence of a large DC current in 
Fig.~\ref{fig:fig_3}.  The dramatic increase in the saturation field 
can also be explained by the competition between the interlayer 
exchange coupling, external, and dipolar fields.

The overall GMR ratio defined as 
$(R_{\mathrm{AP}}-R_\mathrm{P})/R_\mathrm{P}$, where $R_{\mathrm{AP}}$ 
is the highest resistance value in the antiferromagnetic configuration 
and $R_{\mathrm{P}}$ denotes the smallest resistance in the saturated 
state, amounts to 2.6\% at RT and 5.6\% at 4\,K.

A DC current influences the resistance $R$ and, at some critical 
values, the magnetization state of the junction 
(Fig.~\ref{fig:fig_3}).  Positive current corresponds to an electron 
flow from the ``free'' Fe(2) to the ``fixed'' Fe(10) layer.  We 
observe a parabolic background which has been measured previously 
\cite{albert00,grollier01,urazhdin03,krivorotov05} and is usually 
explained by Joule heating of the junction.  On top of that, we 
measure field dependent resistance changes, which can be attributed to 
spin-torque effects.  For instance at -20\,mT 
[Fig.~\ref{fig:fig_3}(b)], the resistance drops at 
$I_\mathrm{C}^+=+18.2\,$mA from the high-resistive to an intermediate 
state.  After reducing the current again, the resistance jumps back to 
the large value.  But also at negative bias the resistance changes at 
$I_\mathrm{C}^-=-12.1\,$mA from large to small.  With an estimated 
junction diameter of $d=150\,$nm the corresponding critical current 
densities are $j_\mathrm{c}^+ = 1\cdot 10^8\,\mathrm{A/cm^2}$ and 
$j_\mathrm{c}^- = -0,7\cdot 10^8\,\mathrm{A/cm^2}$.  As already 
mentioned in the discussion about the GMR data of Fig.~\ref{fig:fig_2} 
we cannot directly relate the resistance jumps to changes between 
specific magnetization states.

If we start the measurement in the intermediate resistance state at a 
field of -31\,mT, the canted alignment is the initial state.  But at 
large positive and negative currents (+31.4 and -24.8\,mA) we observe 
strong deviations from the parabolic background that may indicate 
current-driven high-frequency excitations of the magnetization.

The occurrence of jumps at both polarities of the current at small 
fields is at first glance surprising, but can be explained by taking 
into account that both Fe/Cr and Fe/Ag interfaces contribute and have 
spin scattering asymmetries with opposite signs 
\cite{stiles00,alhajdarwish04}.  This leads for the Fe/Cr subsystem to 
inverse current-induced magnetization switching, very similar to 
inverse GMR \cite{geor94-1,buch03-2}.  Thus, the spin torques for the 
two subsystems are inverted.  For instance, a negative current 
stabilizes the parallel state for Fe/Ag/Fe and the antiparallel state 
for Fe/Cr/Fe.  At low fields, the central Fe(10) layer points opposite 
to the external magnetic field (Fig.~\ref{fig:fig_1}).  At positive 
currents, the spin-transfer torque generated in the Fe/Cr subsystem 
destabilizes this direction and switches the Fe(10) layer 
[Fig.~\ref{fig:fig_3}(b)].  At negative currents, the Fe(2) layer gets 
unstable by the torque created from the Fe/Ag subsystem, while the 
Fe/Cr subsystem is even stronger stabilized in the antiparallel state.
   
At large magnetic fields exceeding the saturation field [\textit{e.g.}, 
-1166\,mT in Fig.~\ref{fig:fig_3}(a)], the two thick bottom layers 
[Fe(14) and Fe(10)] are stronger stabilized by the Zeeman energy than 
the Fe(2) layer, and therefore only one step-like resistance change 
due to magnetic excitations of Fe(2) at $I>0$ is observed under these 
conditions.


In conclusion, we have prepared single-crystalline nanopillars by 
molecular beam epitaxy and a combined process of optical and e-beam 
lithography.  The extended multilayers are characterized by MOKE and 
compared to CPP-GMR data of the nanopillars, which clearly show the 
4-fold magnetocrystalline anisotropy of Fe.  The large GMR ratio of up 
to 5.6\% at 4\,K reflects the high spin polarization predicted in 
Refs.~\cite{stiles00,stiles02}.  After the patterning process the 
magnetic properties change so that the ``free'' Fe(2) layer is now 
coupled to the rest of the system by dipolar stray fields.  Under the 
influence of a DC current we are able to measure distinct resistance 
changes, which give clear evidence of spin-torque effects at current 
densities of about $10^8\,\mathrm{A/cm^2}$.  This value is mostly 
determined by the sizable dipolar coupling in the nanopillar.  At high 
magnetic fields, step-like resistance changes are measured at positive 
currents and are attributed to current-driven magnetic excitations.  
The opposite spin scattering asymmetries of Fe/Cr and Fe/Ag enable us 
to observe CIMS at small magnetic fields for both subsystems in a 
single device.  The switching at opposite current polarity provides 
opportunities for optimizing the CIMS behavior and realizing further 
magnetic excitation dynamics, \textit{e.g.} by exciting one subsystem 
at higher current density while simultaneously suppressing excitations 
of the fixed layer with the torque exerted by the second subsystem.


\end{document}